\documentclass[twocolumn,showpacs,preprintnumbers,amsmath,amssymb]{revtex4}
\usepackage{graphicx}% Include figure files
\usepackage{dcolumn}% Align table columns on decimal point
\usepackage{bm}% bold math
\usepackage[tight]{subfigure}
\usepackage{amsmath}
\usepackage{verbatim}
\usepackage{color}
\newcommand{\half}{\textstyle \frac{1}{2}}

\begin{document}
\title{
Non-linear thermoelectrics of molecular junctions with vibrational coupling
}
\author{M. Leijnse$^{(1)}$}
\author{M. R. Wegewijs$^{(2,3,4)}$}
\author{K. Flensberg$^{(1)}$}
\affiliation{
  (1) Nano-Science Center, Niels Bohr Institute,
      University of Copenhagen,
      2100~Copenhagen \O, Denmark 
      \\
  (2) Institut f\"ur Theoretische Physik A,
      RWTH Aachen, 52056 Aachen,  Germany \\
  (3) Institut f\"ur Festk{\"o}rper-Forschung - Theorie 3,
      Forschungszentrum J{\"u}lich, 52425 J{\"u}lich,  Germany \\
  (4) JARA- Fundamentals of Future Information Technology\\
}
\begin{abstract}
We present a detailed study of the \emph{non-linear} thermoelectric properties of a molecular junction,
represented by a dissipative Anderson-Holstein model.
A single orbital level with strong Coulomb interaction is coupled to a localized vibrational mode
and we account for both electron and phonon exchange with both electrodes,
investigating how these contribute to the heat and charge transport.
We calculate the efficiency and power output of the device operated as a heat to electric power 
converter and identify the optimal operating conditions, which are found to be qualitatively changed by 
the presence of the vibrational mode. 
Based on this study of a generic model system, we discuss the desirable properties of molecular 
junctions for thermoelectric applications.
\end{abstract}
\maketitle
\section{Introduction}
The field of single-molecule electronics has been expanding rapidly during 
recent years, as techniques to electrically contact and control single molecules in a
transport junction have improved~\cite{Reed97, Stipe98, Park99elmig, Kubatkin03, ONeill05}. 
By studying the electric current through the molecule as function of the applied  
voltage-bias, spectroscopic information can be extracted~\cite{Stipe98}. In setups 
with a gate-electrode, which can be used to control the electrostatic 
potential on the molecule, a detailed  
spectroscopy can be performed~\cite{Park99elmig, Kubatkin03, ONeill05}.
By applying a temperature-bias and measuring the induced electric current
or voltage, additional information
can be extracted, such as the type of carriers (holes / electrons) dominating
transport~\cite{Paulsson03}. This emerging field of molecular 
thermoelectrics~\cite{Reddy07, Baheti08, Finch09, Murphy08, Koch04a} is also interesting for 
applications.
Molecules have been predicted to be particularly efficient for 
conversion of heat into electric energy~\cite{Finch09, Murphy08} (or analogously for cooling,
using electric energy to pump heat),
the reason being their very sharp electronic resonances when weakly coupled to electrodes~\cite{Murphy08}.
This is similar to the large thermoelectric efficiency of e.g., 
semi-conducting nanowires with highly peaked densities of states~\cite{Lin00}.
\par
Most theoretical works on meso- and nano-scale thermoelectrics have focused on the 
\emph{linear, equilibrium} regime, where
one operates close to the small voltage $V=-S\Delta T$
which exactly cancels the current induced by the small thermal bias $\Delta T$.
Here the thermopower (or Seebeck coefficient) $S = G_T / G$ is the decisive quantity, where 
($G_T$) $G$ is the (thermal) conductance.
A large efficiency $\eta$ of the device operated as a heat to electric energy converter
is then related to a large dimensionless 
thermoelectric figure of merit $ZT = G S^2 T / \kappa$, where $T$ is the operating temperature 
and $\kappa$ the thermal conductance. 
In bulk systems, $ZT$ is normally limited by the Wiedemann-Franz law, 
stating that $\kappa / G T$ is a system independent constant.
However, the Wiedemann-Franz law is a result of Fermi-liquid theory and breaks down
in mesoscopic and nanoscopic systems, e.g., due to large Coulomb interaction, as has been
demonstrated for quantum dots~\cite{Boese01b} and metallic
islands~\cite{Kubala08}, allowing much larger values of $ZT$ to be reached.
As $ZT \rightarrow \infty$, the efficiency approaches the ideal Carnot value
$\eta \rightarrow \eta_C = 1 - T / (T + \Delta T)$~\cite{Murphy08}.
\par
However, in the linear regime, $\Delta T \ll T$, 
the efficiency stays low even if $ZT$ can be made very large: $\eta \approx \Delta T/T \ll 1$.
The  \emph{non-linear} thermoelectric properties of molecular junctions are 
therefore of great interest.
Recent experiments~\cite{Reddy07} probing the thermopower of thiol end-capped organic molecules 
showed non-linearities in the measured $S$ already at $\Delta T \approx 0.1 T$.
Earlier measurements of thermopower in metallic island single-electron transistors
even displayed a change of the sign of the thermopower for very large $\Delta T$~\cite{Staring93}.
\par
In the interesting regime of sharp electronic resonances, the electron tunnel coupling $\Gamma$ is small and
the main factor limiting the efficiency of molecular energy 
converters is expected to be the heat current from phonon exchange with rate $\gamma$~\cite{Murphy08}.
Nonetheless, to our knowledge, its effect has this far not been systematically 
investigated.
Only by making the tunnel coupling larger, $\Gamma \gg \gamma$, the phonon contribution to 
the heat current becomes negligible. In this case, however,
the efficiency becomes instead limited by the large electronic life-time broadening of
the molecular resonances. 
The thermoelectric efficiency in this limit of coherent transport was studied very recently in the 
non-linear regime~\cite{Bergfield10} using both a many-body transport approach and a 
(non-interacting) approach based on H\"{u}ckel theory.
Except for the latter work and a few others~\cite{Koch04a, Murphy08},
most theoretical studies of molecular thermoelectrics have focused on non-interacting models,
using a Landauer type approach.
However, 
in the regime of weak tunnel coupling between molecule and electrodes, 
intra-molecular interactions typically constitute the largest 
energy scales of the problem.
\par
In this paper, we calculate the thermoelectric efficiency and converted electric power
of a molecular device, including a single dominant molecular orbital, strong Coulomb 
interaction and coupling to a discrete vibrational mode,
as well as coupling to lead phonons and lead electrons.
Importantly, we include on equal footing the phonon and electron contributions to the heat current,
both of which contribute in establishing the stationary occupation of the  molecular vibrational mode.
The Coulomb repulsion and electron-vibration coupling on the molecule are treated non-perturbatively
in the limit of weak electron and phonon exchange in which thermoelectric efficiency is high.
A central finding is that optimal thermoelectric operation typically is achieved
in the \emph{non-linear, non-equilibrium} regime.
Here concepts of figure of merit and thermopower are no longer meaningful
and the molecular occupancies,  efficiency and output power must be explicitly calculated.
The paper is organized as follows: Sect.~\ref{sec:model} introduces the dissipative
Anderson-Holstein model and the thermoelectric transport equations.
In Sect.~\ref{sec:transport} we present results for the efficiency and output power as function 
of the applied bias voltage and energy of the molecular orbital dominating transport. 
The heating of the molecule is analyzed in Sect.~\ref{sec:heat}
and the optimal choice of molecule and junction parameters is discussed in Sect.~\ref{sec:optimal}.
Section~\ref{sec:conclusions} summarizes and provides an outlook.
\par
Throughout the paper we set $\hbar = k_B = e = 1$, where $\hbar$ is Planck's constant,
$k_B$ the Boltzmann constant and $-e$ the electron charge.
\section{Model and thermoelectric transport theory\label{sec:model}}
\begin{figure}[t!]
  \includegraphics[height=0.55\linewidth]{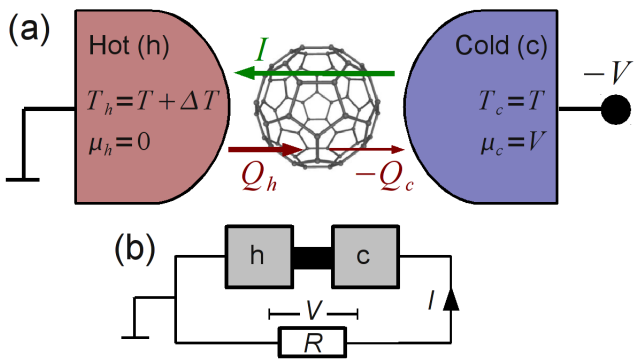}
  \caption{
    \label{fig:1}
    (Color online).
    (a):
    Sketch of a thermoelectric junction with a single molecule, drawn by way of illustration as C$_{60}$.
    Operated as a thermal- to  electric power converter,
    a temperature-bias is applied across the device.
    For an energy level above the electrochemical potentials, $\epsilon > \mu_r$,
    this can drive a net flow of \emph{electrons} from the hot (h) to the cold (c) electrode
    by tunneling through the junctions. 
    In addition, combined electron and phonon heat currents
    $Q_h$ and $Q_c$ are driven through the molecule.
    (b):
    The thermoelectric circuit 
    is loaded by a resistor $R$, and as a result a voltage bias is applied to the cold electrode, 
    partially opposing the thermally induced electron flow.
    The voltage thus ranges from $0$ (corresponding to $R = 0$) to the value where $I = 0$ 
    (corresponding to $R \rightarrow \infty$).
    In the rest of the paper we will however consider a test device as drawn in (a), where $V$,
    rather than $R$, is the free parameter.
    We investigate how to adjust $V$ and the other parameters to obtain a maximal 
    efficiency $\eta$ and output power $P$.
  }
\end{figure}

Despite polarization and screening effects in molecular junctions~\cite{Kaasbjerg08},
the electronic level-spacing in molecular devices is typically large compared to applied 
voltage- and temperature-bias.
We therefore restrict our attention to a single molecular orbital dominating transport.
In fact, the thermoelectric properties have also been predicted to be optimal in this 
case~\cite{Mahan96, Humphrey05}.
However, the quantized vibrational modes,
which couple to the charge localized on a molecular device, 
cannot be neglected~\cite{Park00, Pasupathy04, Osorio07a}.
The vibrations additionally couple to bulk phonon modes of the 
electrodes~\cite{Braig03a}.
The goal of this paper is to clarify the importance of these excitations, 
characteristic of a molecular device, for the thermoelectric properties.
We consider a thermoelectric junction as sketched in Fig.~\ref{fig:1}.
Its basic physics is captured by the following \emph{dissipative} Anderson-Holstein model Hamiltonian
$H_\text{tot} = H + H_R + H_T$, where
\begin{eqnarray}
  \label{eq:3_H}
	H   &=& \tilde{\epsilon} n + \tilde{U} n_\uparrow n_\downarrow
		+ \omega b^{\dagger} b + 
		\omega \lambda n \left( b^\dagger + b \right), \\
  \label{eq:3_HR}
	H_R &=& \sum_{k \sigma r} \epsilon_{k r} 
		c_{k \sigma r}^\dagger c_{k \sigma r} +
		\sum_{q r} \omega_{q r} b_{q r}^{\dagger} b_{q r}, \\ 
  \label{eq:3_HT}
	H_T &=& \sum_{k \sigma r} \tilde{t}_r c_{k \sigma r}^\dagger d_\sigma + h.c. \nonumber \\
	    &+& 	\sum_{q r} C_{q r} \left( b_{q r}^\dagger + b_{q r} \right) 
			\left( b^\dagger + b \right).
\end{eqnarray}
The molecular Hamiltonian, $H$, describes a spin-degenerate orbital level (operator $d_\sigma$ 
for spin-projection $\sigma=\pm \half$) with energy $\tilde{\epsilon}$ 
and Coulomb repulsion $\tilde{U}$.
The electron number $n = \sum_\sigma n_\sigma$, with $n_\sigma = d_\sigma^{\dagger} d_\sigma$,
is linearly coupled to the vibrational coordinate of the harmonic mode of frequency $\omega$ (operator $b$).
The dimensionless electron-vibration coupling $\lambda$ is the shift of the vibrational potential 
as the molecule is charged, measured in units of the vibrational zero-point amplitude.
The reservoir Hamiltonian, $H_R$, describes the combined electron and phonon degrees of freedom in the
two reservoirs, conveniently referred to as the hot ($r = h$) and cold ($r = c$) electrodes.
Non-interacting reservoir electrons 
with energy $\epsilon_{k r}$ are created (annihilated) by 
$c_{k \sigma r}^\dagger$ ($c_{k \sigma r}$); 
$b_{q r}^\dagger$ ($b_{q r}$) are the corresponding 
phonon operators for an electrode phonon mode with frequency $\omega_{q r}$.
In each electrode electrons and phonons are assumed to be in equilibrium with temperatures 
$T_h = T + \Delta T$ and $T_c = T$ respectively.
The coupling between the reservoirs and the molecule is described by the
Hamiltonian $H_T$, where the first term describes tunneling of electrons 
with amplitude $\tilde{t}_r$
and the second term couples the molecular and electrode 
vibrational coordinates with (in general energy-dependent) amplitude $C_{q r}$.
In view of the thermoelectric efficiency we consider the case where both couplings are weak, 
i.e., we want a small tunnel broadening and 
a small heat current carried by the phonons.
Therefore, $H_T$ can be treated perturbatively below.
The electron-electron interaction and electron-vibration coupling on the molecule are 
however allowed to take arbitrary values,
which is a crucial aspect for addressing the important regime $\lambda \sim 1$ and $U \gg T$.
\par
We consider the thermoelectric junction in Fig.~\ref{fig:1}(a) operated as a heat to electric power converter.
One electrode is heated (referred to as hot (h)) with the 
other electrode (referred to as cold (c)) kept at the ambient temperature. 
The hot electrode is grounded (chemical potential $\mu_h = 0$ measured relative to the 
electrode Fermi levels at zero bias) and a (negative) voltage $-V$ is applied to 
the cold electrode ($\mu_c = V > 0$).
For simplicity we assume the capacitances associated with the tunnel junctions to 
both electrodes to be equal,
resulting in a voltage-dependence of the molecular orbital, $\epsilon \propto V/2$.
We note that in an actual device which also makes use of the converted power, the voltage is not applied, 
but rather controlled by the temperature-bias and the resistance of the external circuit,
see Fig.~\ref{fig:1}(b). 
\par
To formulate the transport equations,
the linear coupling term in~(\ref{eq:3_H}) is first eliminated by a standard 
transformation~\cite{LangFirsov63, Flensberg03}, which leads to a renormalization of
the onsite and charging energies:
$\epsilon = \tilde{\epsilon} - \lambda^2 \omega$, $U = \tilde{U} - 2 \lambda^2 \omega$.
After this transformation the eigenstates of $H$ are easily found to 
be given by $|a\rangle |m\rangle$, where 
$a = \{ 0,\uparrow, \downarrow, 2 \}$ is the electronic 
state and $m = \{ 0, 1, 2, \hdots \}$ denotes the vibrational excitation number.
The corresponding eigenenergies are $E_{a m} = E_a + m \omega$ with 
$E_0 = 0$, $E_\sigma = \epsilon$ and $E_2 = 2 \epsilon + U$. 
Furthermore, the electron tunnel amplitude is renormalized to
$t_r = \tilde{t}_r \text{exp}[-\lambda(b^\dagger - b)]$,
thereby incorporating the Franck-Condon factors for electron tunneling.
The resulting transport characteristics under an applied voltage-bias 
have been analyzed in many works, see e.g.,~\cite{Flensberg03, Mitra04b, Leijnse08a}.
We note that in principle the additional coupling to reservoir phonon modes 
requires a more involved transformation, leading to more complicated expressions
for the renormalized parameters~\cite{Braig03a}. 
These corrections can be neglected in the regime considered here, where
the coupling between the reservoir phonons and the molecular vibrational mode is weak.
\par
As mentioned above, the maximum efficiency of energy conversion is expected in the limit of weak 
electron tunneling~\cite{Murphy08} and weak coupling between molecular- and electrode vibrations:
$T \gg \Gamma_r, \gamma_r$.
Here the rate for electron tunneling involving electrode $r$ is $\Gamma_r = 2\pi \rho_r |\tilde{t}_r|^2$, where 
$\rho_r$ is the density of states, which is assumed energy independent (wide band limit).
The relevant rate for phonon exchange with electrode $r$ is
$\gamma_r = 2\pi d_r(\omega) |C_{q_0 r}|^2$, with $q_0$ defined by $\omega_{q_0 r} = \omega$.
Here the phonon density of states, $d_r(E)$, and the coupling strength, $C_{q r}$,
are in general energy-dependent.
However, in the weak coupling limit only their value at $\omega_{q_0 r} = \omega$ enters into the problem
due to the selection rule $|m-m'|=1$ in lowest order perturbation theory
in the coupling to electrode phonons 
(see the expression~(\ref{eq:phonon_rate}) for the rate matrix below).
In the regime of non-linear temperature- and/or voltage-bias addressed in this paper,
the molecular density matrix is not known a priori and needs to be calculated.
For the weak coupling considered here this can be done using a standard master equation approach.
We can neglect contributions from non-diagonal elements of the density matrix,
since the molecular states in our model are non-degenerate on the scale set by the rates 
($\omega \gg \Gamma_r, \gamma_r$) and spin-degeneracy does not lead to off-diagonal contributions.
We note that this holds only in the weak coupling limit where the transport rates are evaluated 
to lowest order perturbation theory in $\Gamma_r$ and $\gamma_r$~\cite{Leijnse08a}.
The transition rates for electron tunneling ($W^{(e)}$) and phonon exchange ($W^{(p)}$) 
can then be calculated from Fermi's golden rule~\cite{Mitra04b, Segal06}:
\begin{widetext}
\begin{eqnarray}\label{eq:electronic_rate}
	W_{am,a'm'}^{(e)} &=& \sum_{\sigma r} \Gamma_r |f_{m m'}|^2  
		\left\{ \delta_{n_a, n_{a'} + 1} \delta_{M_a - M_a',\sigma} 
		f_r(E_{a m} - E_{a'm'}) + 
		\delta_{n_a, n_{a'} - 1} \delta_{M_a' - M_a,\sigma}
		\left[ 1 - f_r(E_{a m} - E_{a'm'}) \right]
		\right\},  \\
                \label{eq:phonon_rate}
	W_{am,a'm'}^{(p)} &=& \sum_{r} \gamma_r \delta_{a a'} \left\{ 
		\delta_{m,m'+1} m b_r(\omega) +
		\delta_{m, m'-1} m' \left[ b_r(\omega) + 1 \right] \right\}.
\end{eqnarray}
\end{widetext}
Here $f_{m m'}$ is a Franck-Condon factor~\cite{Flensberg03}, $M_a$ the 
spin-projection onto the $z$-axis of state $a$, which has electron number $n_a$, and 
$f_r(E) = 1/\{\text{exp}[(E-\mu_r)/T_r] + 1\}$ and $b_r(E) = 1/\{\text{exp}[E/T_r] - 1\}$
are respectively the Fermi- and Bose distribution functions of lead $r$ 
with electro-chemical potential $\mu_r$.
The stationary state master equation to be solved for the occupations $P_{am}$ then reads
\begin{eqnarray}\label{eq:master_equation}
	0 &=& \sum_{a'm'} \left( W_{am,a'm'}^{(e)} + W_{am,a'm'}^{(p)} \right) P_{a'm'}, \\
\label{eq:prob_norm}
	1 &=& \sum_{a m} P_{a m},
\end{eqnarray}
where Eq.~(\ref{eq:prob_norm}) expresses probability normalization.
At this point, we note that the interplay of charge and phonon 
tunneling is still non-trivial, as they do "interact" via the vibrational occupation number.
A finite electric current tends to highly excite the vibrational mode, 
leading to high effective molecular temperatures (see also Fig.~\ref{fig:3}(c)) and 
even clear deviations from an equilibrium (Boltzmann) shape of the distribution.
This effect is particularly pronounced when the electron-vibration coupling is not too 
large, $\lambda \lesssim 1$.
The phonon current, on the other hand, tends to thermalize the vibration towards a temperature, that 
depends only on the temperatures of the hot ($T_h$) and cold ($T_c$) lead and the relative 
size of the couplings $\gamma_h$ and $\gamma_c$. However, through the excitations created by the 
electric current, the phonon current acquires an indirect dependence on both voltage and level position.
Therefore accurate calculation of the non-equilibrium molecular state accounting 
for both electron and phonon effects is crucial.
\par
The electric current, $I_r$, and heat current, $Q_r$, going out of lead $r$, are given by
\begin{eqnarray}
\label{eq:electric_current}
	I_r &=& - \left \langle -\sum_{k \sigma} \frac{d N_{k \sigma r}}{dt} \right \rangle \\
	    &=& - \sum_{a m}\sum_{a' m'} \left( W_{I_r}^{(e)}\right)_{am,a'm'} P_{a' m'}, \\
\label{eq:heat_current}
	Q_r &=& \left \langle -\sum_{k \sigma} (\epsilon_{k r} - \mu_r) 
		\frac{d N_{k \sigma r}}{dt} - 
		\sum_{q} \omega_{q r} \frac{d M_{q r}}{dt} \right \rangle \\
	    &=& \sum_{a m} \sum_{a' m'} \left[ \left( W_{Q_r}^{(e)}\right)_{am,a'm'} +
		\left( W_{Q_r}^{(p)}\right)_{am,a'm'} \right]  P_{a' m'}, \nonumber \\
\end{eqnarray} 
where $N_{k \sigma r} = c_{k \sigma r}^{\dagger} c_{k \sigma r}$ and
$M_{q r} = b_{q r}^{\dagger} b_{q r}$. 
The electron current matrix $W_{I_r}^{(e)}$ is
similar to~(\ref{eq:electronic_rate}), but includes only processes involving
reservoir $r$ and a plus / minus sign for processes adding electrons 
to the molecules (first term in~(\ref{eq:electronic_rate})) / 
removing electrons from the molecules (second term 
in~(\ref{eq:electronic_rate})). Analogously, the heat current matrices, 
$W_{Q_r}^{(e)}$ and $W_{Q_r}^{(p)}$, are similar to~(\ref{eq:electronic_rate})
and~(\ref{eq:phonon_rate}), respectively, but including only processes involving
reservoir $r$ and with the rate multiplied by the energy of the tunneling electron 
(measured relative to $\mu_r$) or phonon.
We note that there is in general some ambiguity associated with the definition of the 
heat current, see~\cite{Wu09}, which however does not matter in the weak coupling 
limit discussed here.
\par
Beyond the linear regime, thermopower and figure of merit are no longer 
suitable quantities and we instead directly calculate the efficiency 
of the energy converter as follows.
Driven by the thermal bias, electrons can gain 
potential energy by tunneling from the hot to the cold electrode via the molecule. 
The resulting electric output power is $P = I V$, where $I = -I_h = I_c$. 
The input heat power, required to maintain the temperature bias, is equal to the 
heat current, $Q_h$, flowing out of the hot electrode. The efficiency is thus given by
\begin{eqnarray}
  \eta &=& I V / Q_h .
\end{eqnarray}
Note that there is no conservation of the stationary heat current,
as there is for the electric current, $I_h + I_c=0$. Instead
the first law of thermodynamics guarantees that $P = Q_h + Q_c$.

\section{Optimal bias voltage and level position\label{sec:transport}}
\begin{figure}[t!]
     \includegraphics[height=0.7\linewidth]{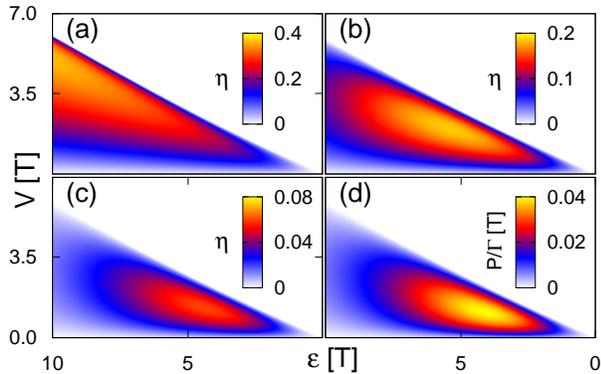}
    \caption{\label{fig:2}
(Color online). 
(a)--(c): Efficiency $\eta$ at thermal bias $\Delta T = T$, as function of  
voltage-bias $V$ and level position $\epsilon$ for increasing coupling to substrate phonons, 
$\gamma = 0$ (a), $\gamma = \Gamma / 10$ (b), $\gamma = \Gamma$ (c). In all plots
$\lambda = 1$, $\omega = T$, $U = 10 T$ and the couplings are symmetric, 
$\gamma_h = \gamma_c = \gamma$, $\Gamma_h = \Gamma_c = \Gamma$.
(d): Output power $P$ as function of $V$ and $\epsilon$ for the parameters
used in (b) (the power depends only weakly on $\gamma$).}
\end{figure}

We start by studying the efficiency and output power at fixed thermal bias, here chosen 
to be $\Delta T = T$, as function of applied voltage-bias $V$ 
and level position $\epsilon$.
The efficiency of a single level quantum dot (spin-less electrons and no vibrational mode)
was studied in~\cite{Esposito09}, where it was shown that the ideal Carnot efficiency is reached 
in the equilibrium limit of vanishing current, requiring the Fermi functions to be equal,
$f_h(\epsilon) = f_c(\epsilon)$, defining a line in the $(V, \epsilon)$ plane:
$V=\epsilon (T_h-T_c)/T_h$.
In Fig.~\ref{fig:2} this equilibrium line corresponds to the boundary of the white areas. 
However, along this line also the output power vanishes 
(corresponding to reversible, infinitely slow operation without entropy loss).
For vanishing couplings to the phonon 
mode, $\gamma_r \rightarrow 0$ and $\lambda \rightarrow 0$, we recover this result in the 
non-interacting limit, $U = 0$, as well as for very strong interactions, $U \gg T, \Delta T$.
In the intermediate regime, the efficiency is slightly reduced.
\par 
Switching on the electron-vibration coupling, but keeping $\gamma_r = 0$, the efficiency
is decreased and never reaches the ideal value ($\eta_C = 0.5$ for $\Delta T = T$), see 
Fig.~\ref{fig:2}(a).
In fact, $\eta$ vanishes close to the zero electric current line (boundary of the white area), 
the reason being that, in contrast to the single-level discussed above, the heat 
current does not vanish completely when the charge current does.
Inside the white area the current has been reversed by a too large voltage-bias and flows from 
high- to low-biased electrode and therefore does not accomplish any useful electric 
work (note that this regime can not be reached in the thermoelectric circuit of Fig.~\ref{fig:1}(b)).
The maximal efficiency is reached in the non-linear regime
when the level is far above the Fermi edges of both leads.
In this case electron transport involves very few thermally excited states 
in the heated electrode (tail of the Fermi function) and electron-induced 
vibrational excitations are exponentially suppressed, minimizing 
electronic heat loss.
However, in this regime the current is highly suppressed, leading to a very small 
output power, see Fig.~\ref{fig:2}(d).
Additionally, even a small coupling to the substrate phonons, 
$\gamma = \Gamma / 10$ in Fig.~\ref{fig:2}(b), drastically decreases the efficiency 
in this low-current regime, while having a much smaller effect in the regime 
where the current is larger ($\epsilon$ is smaller). 
Thus, already a weak coupling to substrate phonon modes, $\gamma \ll \Gamma$, drastically changes the ideal 
operating conditions for maximum efficiency by introducing a heat loss 
which depends only weakly on $\epsilon$ and $V$
(the dependence is indirect, through the vibrational occupations).
When the coupling to the substrate phonons becomes 
comparable to the tunnel coupling, $\gamma \approx \Gamma$ in Fig.~\ref{fig:2}(c), the 
efficiency is significantly decreased also in the high current regime.
The output power, shown in Fig.~\ref{fig:2}(d) for the parameters used in (b), 
depends only weakly on $\gamma$ and is maximal for 
$\epsilon \approx 4 T$ and $V \approx T$.
\section{Temperature dependence and molecular heating\label{sec:heat}}
\begin{figure}[t!]
  \includegraphics[height=0.7\linewidth]{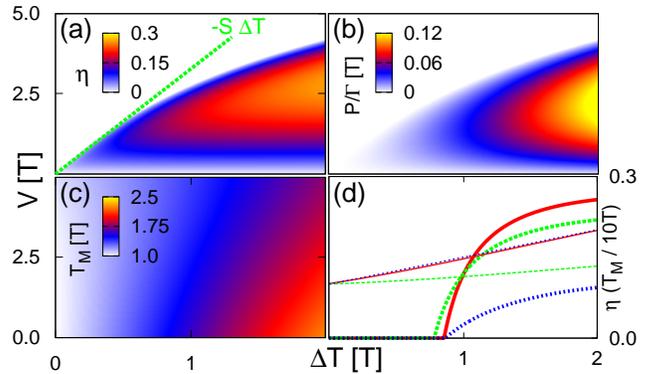}
  \caption{\label{fig:3}
    (Color online). 
    Efficiency $\eta$ (a), output power $P$ (b), and molecular temperature $T_M$ (c),  
    as function of $V$ and $\Delta T$, with $\epsilon = 5 T$ and other parameters as 
    in Fig.~\ref{fig:2}(b).
    Along the edge of the white areas in (a) and (b) $I = 0$, and this edge therefore defines the non-linear 
    thermopower $S(\Delta T) = -V / \Delta T$ at $I = 0$. The standard linear response thermopower,
    $S = S(\Delta T \rightarrow 0)$, is thus given by the slope close to zero, indicated by 
    the green dashed line in (a).    
    Note that $T_M$ in (c) is plotted also in the regime where the current has been reversed, 
    even though the device would normally not be operated under such conditions.
    (d): $\eta$ (thick lines) and $T_M$ (thin lines, normalized by $10T$) as function of 
    $\Delta T$ at $V = 2.5 T$ for 
    $\gamma_h = \gamma_c = \Gamma / 10$ (red solid lines), 
    $10 \gamma_h = \gamma_c = \Gamma$ (green dashed lines), and
    $\gamma_h = \gamma_c = \Gamma$ (blue dotted lines).
}
\end{figure}

Next we fix the level position to a value with both large power and efficiency,
$\epsilon = 5 T$, and vary instead $V$ and $\Delta T$. 
The resulting efficiency and output power is shown in 
Fig.~\ref{fig:3}(a) and (b) respectively, for the same parameters as in 
Fig.~\ref{fig:2}(b).
As above, a too large voltage-bias compared to the temperature-bias reverses the current and 
no useful electric work is accomplished (white areas).
The non-linear thermopower can be defined through $V = -S(\Delta T) \Delta T$ at $I = 0$, i.e., 
$V$ is the finite voltage needed to compensate the temperature-bias and give zero electric current. 
Thus, $-S(\Delta T)$ is given by the slope of 
the line which passes through zero voltage at zero temperature-bias and hits the edge of the 
white areas at $\Delta T$ in Fig.~\ref{fig:3}(a) and (b).
For large temperature-bias there are clear deviations from the linear response thermopower,
$S = S(\Delta T \rightarrow 0)$, given by the slope of the green dashed line in Fig.~\ref{fig:3}(a).
\par
As expected, the efficiency and (even more so)
the power is increased by an increased temperature-bias. 
Figure~\ref{fig:3}(a)--(b) shows the dependence on the temperature-bias over a wide range, 
all the way up to $T_h = 3T$. Such large temperature-bias could be obtained if the device is
operated at low temperatures. 
In applications, however, $T$ is most likely room temperature and junction stability
limits operation to lower relative temperature-bias (e.g., $\Delta T = T/3$ would mean 
$T_h \approx 400$~K, which is a realistic value).
However, it is actually possible to keep the (non-equilibrium) molecular temperature, $T_M$,
much lower than the average electrode temperature, $T_M < \bar T = T +\Delta T/2$, 
allowing operation at higher temperature-bias. 
\par
To calculate $T_M$ we use
an idea suggested in~\cite{Galperin07b} and couple an additional phonon bath
("thermometer") very weakly to the molecule.
The temperature of the thermometer bath is varied and
$T_M$ is defined as the bath temperature where the heat current
between this bath and the molecule vanishes.
Figure~\ref{fig:3}(c) shows $T_M$ as function of $V$ and $\Delta T$, where it is seen that 
for small voltages $T_M$ exceeds the average electrode temperature $T_M > \bar T$.
A larger voltage-bias, however, reduces the non-equilibrium  electron current and $T_M$ approaches $\bar T$. 
Still, it is desirable to reduce $T_M$ further, thereby allowing operation at higher $\Delta T$
without breaking the molecule.
In designing molecular thermoelectric junctions it is therefore important to
choose the electrode material and molecular anchoring groups such that 
the molecular vibration couples more strongly to the substrate phonons 
of the colder electrode ($\gamma_c \gg \gamma_h$).
This is shown in Fig.~\ref{fig:3}(d), where 
$\gamma_c = \gamma_h = \Gamma / 10$ (red solid lines),
$\gamma_c = 10 \gamma_h = \Gamma$ (green dashed lines) and
$\gamma_c = \gamma_h = \Gamma$ (blue dotted lines).
The asymmetric phonon coupling significantly reduces $T_M$ (thin lines)
by preventing phonons from accumulating on the molecule: they enter slowly ($\gamma_h$) and exit quickly ($\gamma_c$).
However, the asymmetry has a rather small effect on the efficiency (thick lines) since heat is still prevented from 
"leaking" through the molecule via the phonons as long as $\gamma_h$ stays small compared to $\Gamma$.
In contrast, with also $\gamma_h$ large the efficiency goes down 
much more, and $T_M$ goes up (the red solid and blue dotted lines for $T_M$ still show 
a very small difference: the electron tunneling and substrate phonon couplings 
drive the system toward slightly different molecular temperatures, causing $T_M$ 
to depend on the ratio $\gamma / \Gamma$).
In realizing the mechanical coupling asymmetry $\gamma_c \gg \gamma_h$, it is important keep 
the electronic tunnel coupling symmetric.
Introducing an asymmetry in the electron tunnel couplings, $\Gamma_h \neq \Gamma_c$, while keeping 
$\Gamma_h + \Gamma_c$ fixed, reduces the efficiency since the current level is set by 
the smallest coupling, while the phonon leakage current depends only weakly on
$\Gamma_h$ and $\Gamma_c$ (only indirectly through the vibrational occupations).
\section{Optimal thermoelectric junctions\label{sec:optimal}}
\begin{figure}[t!]
  \includegraphics[height=0.7\linewidth]{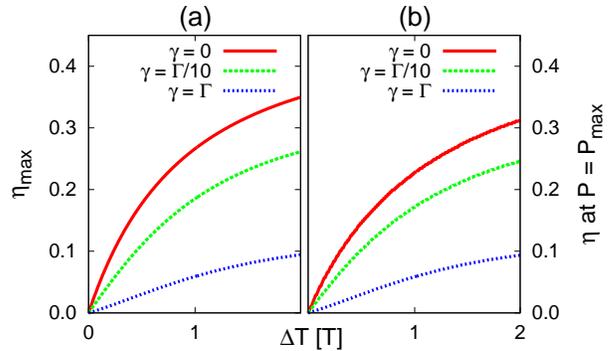}
  \caption{\label{fig:4}
    (Color online). 
    (a): 
    Maximum efficiency, $\eta_\text{max}$, as function of $\Delta T$, i.e., 
    the maximum obtained from vertical cuts in a plot such as Fig.~\ref{fig:3}(a).
    Parameters as in Fig.~\ref{fig:3}(a)--(c), 
    but the strength of the coupling to substrate phonons is varied. Here 
    $\gamma = \gamma_h = \gamma_c$ and $\Gamma = \Gamma_h = \Gamma_c$.
    (b):  
    Same as (a), but instead showing efficiency at maximum output power $P_\text{max}$, 
    i.e., $\eta$ taken at the maximum $P$ obtained from vertical cuts in a plot such as 
    Fig.~\ref{fig:3}(b).}
\end{figure}

Comparing Fig.~\ref{fig:3}(a) and (b), we reach an important result for optimizing 
molecular thermoelectric junctions,
namely that for a given temperature-bias, 
maximum efficiency and maximum output power is achieved at almost the same voltage-bias.
This is also seen in Fig.~\ref{fig:4},
where (a) shows the maximum efficiency, $\eta_\text{max}$,
and (b) the efficiency $\eta$ at maximum power $P = P_\text{max}$,
both obtained by adjusting the voltage at given thermal bias $\Delta T$.
For non-zero coupling $\gamma$ to substrate phonons the 
maximum efficiency and the efficiency at maximum output power are very close.
The reason is seen from the relation $P=\eta~( Q_h^{(e)} + Q_h^{(p)}) $,
where $Q_h^{(e)}$ ($Q_h^{(p)}$) is the electron (phonon) contribution to the heat current.
Since $Q_h^{(p)}$ only has a weak (indirect) dependence on the voltage-bias,
$\eta$ and $P$ can be simultaneously maximized by adjusting the bias when the phonon heat loss 
dominates ($\gamma \gg \Gamma$). As Fig.~\ref{fig:3} and Fig.~\ref{fig:4} shows, this holds approximately 
also when $\gamma \lesssim \Gamma$.
\par
\begin{figure}[t!]
  \includegraphics[height=0.7\linewidth]{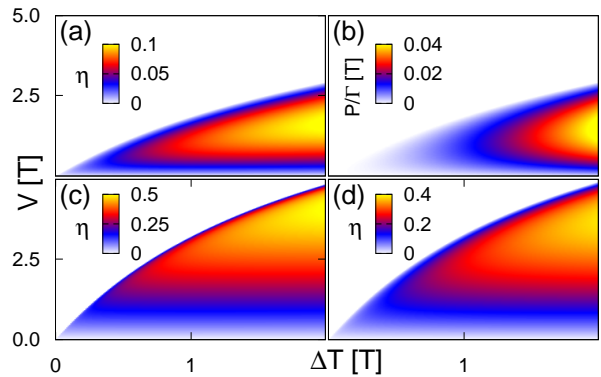}
  \caption{\label{fig:5}
    (Color online).
    Dependence on the molecular vibration frequency $\omega$ 
    and its coupling to the electron charge $\lambda$,
    all other parameters fixed to those of Fig.~\ref{fig:3}(a)--(c).
    (a)--(b): Efficiency $\eta$ (a) and power $P$ (b),   
    as function of $V$ and $\Delta T$ for $\lambda = 2$ and $\omega=T$.
    (c)--(d): Efficiency $\eta$ as function of $V$ and $\Delta T$
    for $\lambda = 0.5$ and $\omega=T$ (c)
    and $\lambda = 1$ and $\omega = T / 10$ (d).
  }
\end{figure}

When the electron-vibration coupling becomes strong, $\lambda > 1$, the tunnel amplitudes involving the 
vibrational ground state become suppressed 
(Franck-Condon blockade~\cite{Koch04b}). This reduces both the efficiency and output power
since the current is decreased. 
Additionally, heat dissipation is increased since transport through excited vibrational states is favored,
 the typical energy transferred to the vibrational mode by a tunneling electron being 
given by the classical displacement energy, $\omega \lambda^2$.
This is shown in Fig.~\ref{fig:5}(a) and (b), where $\lambda = 2$ 
(cf., Fig.~\ref{fig:3}(a) and (b)).
Similarly, a smaller electron-vibration coupling enhances the efficiency, see 
Fig.~\ref{fig:5}(c), where $\lambda = 0.5$.
\par
In choosing $\omega = T$ ($\approx 25$ meV assuming room temperature) we have 
investigated the influence of a rather high 
energy vibrational mode.
Molecules often also have vibrational modes with much lower frequency, down to a few meV,
especially when contacted to electrodes by linker wires~\cite{Seldenthuis08}.
However, as is shown in Fig.~\ref{fig:5}(d), where $\omega = T / 10$, a low-energy
vibrational mode leads to a much smaller decrease of $\eta$ compared to the ideal 
case of no vibrational mode, as long as $\lambda$ is not too large.
The reason is simply that a low frequency mode essentially can be seen as a broadening of the 
electronic resonance of width $\sim \omega \lambda^2$, setting the scale for additional heat loss 
from electron tunneling compared to the case of no vibrational mode.
Almost all decrease in efficiency in this case comes from the coupling to substrate phonon 
modes. 
In contrast, a vibrational mode with a frequency much larger than the involved 
temperature- and voltage-bias, does not contribute at all to electron or heat transport 
(other than through the trivial shift of the electronic parameters 
$\tilde{\epsilon}$, $\tilde{U}$ and $\tilde{t_r}$ through the electron-vibration
coupling).
\par
Finally, we mention that in the simple model analyzed here, the strength of the Coulomb interaction
does not play a crucial role. In the presence of a coupling to substrate phonons, reducing $U$ 
leads to a somewhat larger efficiency (and output power) as the electric current is increased by 
the presence of another "transport channel".
\section{Conclusions\label{sec:conclusions}}
We have analyzed the efficiency and output power of a non-linear molecular thermoelectric device
operated as a power converter.
Accounting for the molecular vibration and its coupling to substrate phonons turned out to be 
crucial in comparison with results for quantum dot models without these,
as it qualitatively changes the operating conditions for optimal efficiency 
away from the equilibrium regime.
By investigating a generic model system we can now identify some basic criteria
for efficient molecular energy converters:
\par
{\it (i)} The coupling between substrate phonon modes and molecular vibrations should be asymmetric and minimal. 
Coupling more strongly to the colder lead reduces the molecular temperature and allows operation at higher 
temperature-bias, improving efficiency and output power.
\par
{\it (ii)} The electron tunnel couplings should be symmetric, $\Gamma_h \approx \Gamma_c$.
Furthermore, they should be small as to minimize the life-time broadening, but still larger 
than the phonon coupling, $\Gamma > \gamma$.
\par
{\it (iii)} The local electron-vibration coupling energy should be small compared to the zero-point 
energy of the vibrational mode ($\lambda < 1$). This is most crucial for vibrational modes with frequencies around 
the operating temperatures and voltages. Modes with much higher frequencies do not contribute at all, 
and those with much lower frequencies only contribute to the heat loss through the coupling to substrate 
phonons.
\par
{\it (iv)} Ideal operating conditions for high efficiency and power is achieved when the conducting 
orbital energy is at a few $k_B T$ from the Fermi edges of the electrodes ($\sim 100$~meV at room temperature).
Control of the thermopower by adding electron donating or withdrawing groups to 
benzenedithiol molecules, thereby shifting the position of the HOMO and LUMO, 
was recently demonstrated~\cite{Baheti08}.
The temperature-bias should be chosen as high as is allowed by molecular stability
and the heat source.
The ideal voltage-bias depends
on the other parameters, but is nearly the same when optimizing output power as when
optimizing efficiency. Additionally, the efficiency at maximum power is very close to the
maximum efficiency.
\par
The general insights obtained in this exhaustive study of the most basic molecular thermoelectric model can serve as a guide
for more complex molecular modeling, incorporating multiple vibrational modes, multiple electronic states, 
breakdown of the Born-Oppenheimer picture (pseudo Jahn-Teller mixing~\cite{Reckermann08b}), etc.
In general, one expects deviations from a single-orbital model to give a 
less efficient energy-converter, as additional heat is lost by population of excited states.
It might however be possible to find special circumstances under which excited states 
can instead be desirable, e.g., by effectively cooling the vibrational mode.
Atomistic studies of specific configurations of molecules, anchoring groups and electrodes
may identify suitable systems which satisfy the above criteria and thereby further assist in
advancing the chemical engineering of molecular thermoelectric junctions.
For device applications, engineering of molecular monolayer devices,
rather than ones based on a single molecule, presents a challenge to supramolecular chemistry,
nanodevice fabrication and surface science.
\section*{ACKNOWLEDGMENTS}
We acknowledge financial support from the DFG under Contract No. SPP-1243 (M.~L., M.~W)  
and the European Union under the FP7 STREP program SINGLE (M.~L., K.~F.).
This work was carried out partly in the Danish-Chinese Centre for Molecular 
Nano-Electronics supported by the Danish National Research Foundation.
\bibliographystyle{apsrev}

\end{document}